\documentclass[final,3p,times,onecolumn]{elsarticle}
\usepackage{graphicx}
\usepackage{amsmath,bm}
\allowdisplaybreaks 
\def\bea#1\eea{\begin{align}#1\end{align}} 
\newcommand{\nnu}{\nonumber\\}
\newcommand{\bef}{\begin{figure}[htb]\centering}
\newcommand{\eef}{\end{figure}}
\usepackage{color}
\usepackage[normalem]{ulem}

\begin{document}

\begin{frontmatter}
\title{Jet fragmentation functions for $Z$-tagged jets}
\author{Zhong-Bo Kang\fnref{label1,label2}}
\ead{zkang@physics.ucla.edu}
\author{Kyle Lee\fnref{label3,label4}}
\ead{kunsu.lee@stonybrook.edu}
\author{John Terry\fnref{label1,label2}}
\ead{johndterry@physics.ucla.edu}
\author{Hongxi Xing\fnref{label5,label6}}
\ead{hxing@m.scnu.edu.cn}
\address[label1]{Department of Physics and Astronomy, University of California, Los Angeles, California 90095, USA}
\address[label2]{Mani L. Bhaumik Institute for Theoretical Physics, University of California, Los Angeles, California 90095, USA}
\address[label3]{C.N. Yang Institute for Theoretical Physics, Stony Brook University, Stony Brook, New York 11794, USA}
\address[label4]{Department of Physics and Astronomy, Stony Brook University, Stony Brook, New York 11794, USA}
\address[label5]{Institute of Quantum Matter, South China Normal University, Guangzhou 510006, China}
\address[label6]{School of Physics and Telecommunication Engineering, South China Normal University, Guangzhou 510006, China}

\begin{abstract}
Recently the LHCb collaboration has measured both longitudinal and transverse momentum distribution of hadrons produced inside $Z$-tagged jets in proton-proton collisions at the Large Hadron Collider. These distributions are commonly referred to as jet fragmentation functions and are characterized by the longitudinal momentum fraction $z_h$ of the jet carried by the hadron and the transverse momentum $j_\perp$ with respect to the jet direction. We derive a QCD formalism within Soft-Collinear Effective Theory to describe these distributions and find that the $z_h$-dependence provides information on standard collinear fragmentation functions, while $j_\perp$-dependence probes transverse momentum dependent (TMD) fragmentation functions. We perform theoretical calculations and compare our results with the LHCb data. We find good agreement for the intermediate $z_h$ region. For $j_\perp$-dependence, we suggest binning in both $z_h$ and $j_\perp$, which would lead to a more direct probing of TMD fragmentation functions. 
\end{abstract}

\begin{keyword}
jets \sep fragmentation functions \sep perturbative QCD \sep Soft Collinear Effective Theory
\end{keyword}

\end{frontmatter}

\section{Introduction}
The momentum distribution of hadrons inside a fully reconstructed jet, commonly referred to as jet fragmentation function (JFF), has received increasing attention in recent years. The JFF probes the parton-to-hadron fragmentation function at a differential level and can thus provide new insights for the hadronization process. Jet fragmentation functions have been measured for single inclusive jet produced in unpolarized proton-proton collisions at the Large Hadron Collider (LHC) for light hadrons~\cite{Aad:2011sc,Aaboud:2017tke}, for open heavy flavor mesons~\cite{Aad:2011td,CMS:2018ovh,Acharya:2019zup}, and for heavy quarkonium~\cite{Aaij:2017fak,CMS:2018mjn}. Such measurements have already started to constrain the fragmentation functions for open heavy flavor mesons~\cite{Chien:2015ctp,Anderle:2017cgl}, and to pin down non-relativistic QCD (NRQCD) long-distance matrix elements, which characterize the hadronization process for heavy quarkonium production~\cite{Kang:2017yde,Bain:2017wvk}. 

The same measurements in heavy ion collisions show a strong modification of the JFF~\cite{Chatrchyan:2012gw,Aaboud:2019oac} in the existence of the hot and dense medium, the quark-gluon plasma, and thus serve as a novel probe for the medium. Jet fragmentation functions can also be measured in transversely polarized proton-proton collisions. For example, the measurements by the STAR collaboration at the Relativistic Heavy Ion Collider (RHIC) study the azimuthal distribution of hadrons inside the jet~\cite{Adamczyk:2017wld} and provide information for the so-called Collins fragmentation functions~\cite{Kang:2017btw,Collins:1992kk,Yuan:2007nd}.

Single inclusive jet production at the LHC involves a large fraction of gluon jets~\cite{Kang:2016ehg}. In order to further disentangle quark and gluon jets, one can study e.g., photon-tagged jet production and the JFF in photon-tagged jets. These processes are more sensitive to the quark jets, or quark-to-hadron fragmentation functions. See~\cite{Aaboud:2019oac} for recent JFF measurement for photon-tagged jets. More recently the LHCb collaboration at the LHC has measured both longitudinal and transverse momentum distribution of charged hadrons produced inside $Z$-tagged jets in the forward rapidity region in proton-proton collisions, $p+p\to Z+{\rm jet}+X$. Experimental requirements are placed on the $Z$-jet pair to better identify events that correspond to a two-to-two partonic hard scattering process, i.e. the $Z$-jet pair is required to be nearly back-to-back in azimuth such that $\left|\Delta \phi_{Z-\rm jet}\right| > 7\pi/8$. In our previous work~\cite{Buffing:2018ggv}, we developed a factorized framework for back-to-back photon-jet production within Soft-Collinear Effective Theory (SCET)~\cite{Bauer:2000ew,Bauer:2000yr,Bauer:2001ct,Bauer:2001yt,Bauer:2002nz}. Such a framework can be generalized to study back-to-back $Z$-jet production~\cite{Chien:2019gyf}, as well as JFF in $Z$-tagged jets. 

In this paper, we derive such a formalism, perform theoretical calculations and compare our results with the LHCb data. The JFFs are characterized by the longitudinal momentum fraction $z_h$ of the jet carried by the hadron and the transverse momentum $j_\perp$ with respect to the jet direction. We demonstrate how the $z_h$-dependence is connected to the standard collinear fragmentation functions, while the $j_\perp$-dependence is associated with the transverse momentum dependent (TMD) fragmentation functions. For the phenomenology, we find good agreement for the intermediate $z_h$ region. For $j_\perp$-dependence, we suggest binning in both $z_h$ and $j_\perp$, which would lead to a more direct probing of TMD fragmentation functions. The rest of the paper is organized as follows. In Sec.~\ref{sec:theory}, we generalize our QCD formalism developed for photon-jet production to describe back-to-back $Z$-jet cross section, as well as the jet fragmentation functions in $Z$-tagged jets. Numerical results are presented in Sec.~\ref{sec:pheno}, where we compare our calculations with the LHCb experimental data. We conclude our paper in Sec.~\ref{sec:summary}.

\section{Theoretical framework}
\label{sec:theory}
We consider hadron distribution inside $Z$-tagged jets in proton-proton collisions, as illustrated in Fig.~\ref{fig:Z-JFF},
\bea
p(p_A)+p(p_B)\to Z(\eta_Z, {\bm p}_{ZT})+{\rm jet}(\eta_J,~{\bm p}_{JT}, R)~h(z_h, {\bm j}_\perp)+X\,,
\eea
where $s=(p_A+p_B)^2$ is the center-of-mass energy squared, the $Z$-boson is produced with the rapidity $\eta_Z$ and transverse momentum ${\bm p}_{ZT}$, while the jet is reconstructed in the usual anti-$k_T$ algorithm~\cite{Cacciari:2008gp} with the jet radius parameter $R$, and the jet has the rapidity $\eta_J$ and the transverse momentum ${\bm p}_{JT}$. One further observes a hadron inside the jet, which carries a longitudinal momentum fraction $z_h$ of the jet, and a transverse momentum ${\bm j}_\perp$ with respect to the jet direction. 

\bef
\includegraphics[width=0.45\columnwidth]{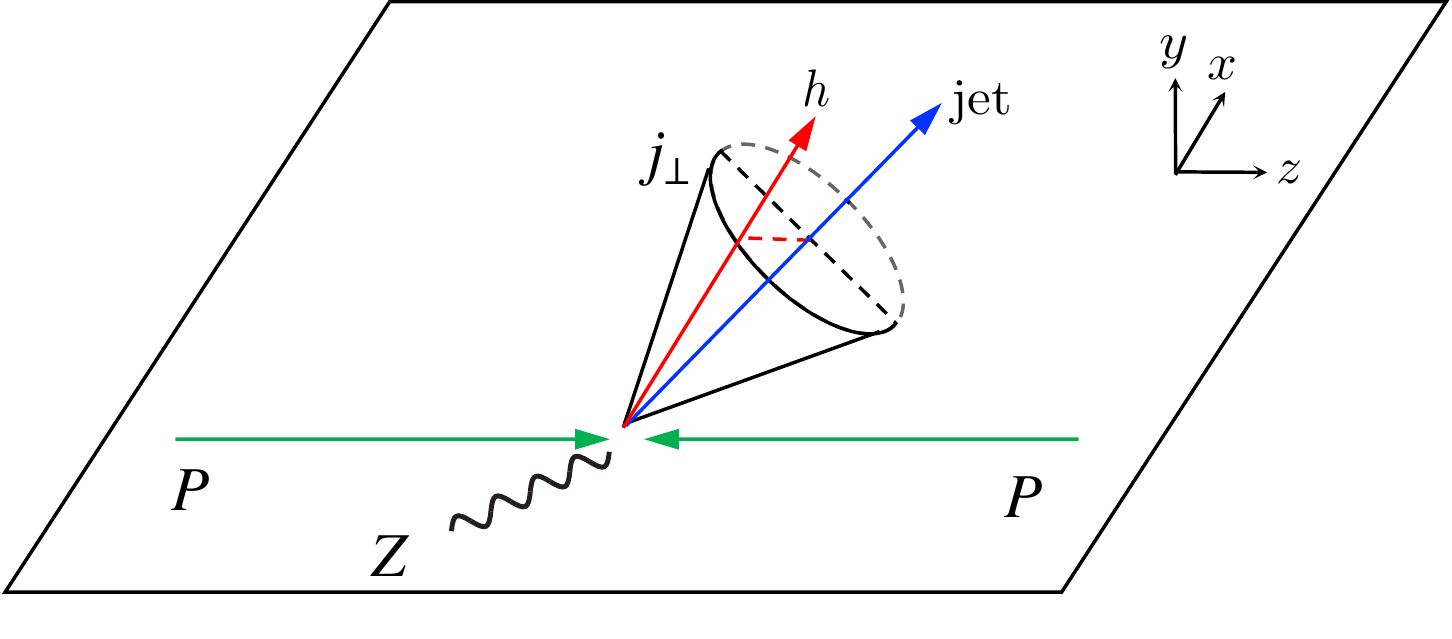}
\caption{Illustration for the distribution of hadrons inside jets in $Z$-tagged jet production in proton-proton collisions.}
\label{fig:Z-JFF}
\eef

One usually defines the imbalance ${\bm q}_T$ between the transverse momenta of the $Z$-boson and the jet, and the average of the transverse momenta ${\bm p}_T$ as
\bea
{\bm q}_T \equiv {\bm p}_{ZT}+ {\bm p}_{JT}, 
\qquad
{\bm p}_T =\frac{{\bm p}_{ZT} - {\bm p}_{JT}}{2}. 
\eea
To be consistent with the experimental setup~\cite{Aaij:2019ctd}, we only consider the region where the $Z$-boson and the jet are produced back-to-back. In such a region, the imbalance is much smaller than the average transverse momentum, $q_T\ll p_T$, where the perturbative computations receive contributions of large logarithms of the form $\alpha_s^n\ln^{2n}(p_T/q_T)$, which have to be resummed. In the following, we first review the QCD formalism that achieves this purpose. We then generalize to the case of hadron distribution inside the jets, for both longitudinal $z_h$-distribution and the transverse momentum $j_\perp$-distribution.

\subsection{$Z$-tagged jet cross section}
A formalism has been developed to resum the logarithms of the form $\alpha_s^n\ln^{2n}(p_T/q_T)$ as well as the logarithms of jet radius $\ln R$ in our previous work~\cite{Buffing:2018ggv} for back-to-back photon-tagged jet cross section. This formalism can be generalized to the $Z$-tagged jet production, $p+p\to Z+{\rm jet}+X$. In such a formalism, the differential cross section  can be written as
\bea
\frac{d\sigma}{d{\cal PS} } = & 
\sum_{a,b,c} \int d \phi_J \, 
\int \prod_{i=1}^4 d^2{\bm k}_{iT}  \delta^{2}({\bm q}_T - \sum_i^4 {\bm k}_{iT})
 f_a(x_a,k_{1T}^2, \mu, \nu) f_b(x_b,k_{2T}^2, \mu, \nu) 
 \nonumber \\
 &
  \times   
  S^{\text{global}}_{n{\bar n} n_J }({\bm k}_{3T}, \mu, \nu) 
  S^{cs}_{n_J}({\bm k}_{4T},R, \mu) 
  H_{a b\rightarrow c Z}(p_T, m_Z, \mu)  \, J_{c}(p_{JT} R, \mu) \,,
\label{eq:jet}  
\eea
where the phase space $d{\cal PS} = d\eta_J d\eta_Z dp_T d^2 {\bm q}_{T}$, and $\phi_J$ is the azimuthal angle of the jet. Besides different hard functions $H_{a b\rightarrow c Z}$, the above formalism is the same as that for photon-tagged jet production developed in~\cite{Buffing:2018ggv}. See also Ref.~\cite{Chien:2019gyf}, where the authors further study the impact of the so-called non-global logarithms~\cite{Dasgupta:2001sh}. 

We include both partonic channels $q\bar q\to gZ$ and $qg\to qZ$ at the next-to-leading order (NLO) for the hard functions $H_{a b\rightarrow c Z}$~\cite{Becher:2011fc,Arnold:1988dp}. On the other hand, $f_a(x_a,k_{1T}^2, \mu, \nu)$ and $f_b(x_b, k_{2T}^2, \mu, \nu)$ in Eq.~\eqref{eq:jet} are the TMD parton distribution functions (PDFs) of parton flavors $a$ and $b$~\cite{Collins:2011zzd}. These TMD PDFs contain so-called rapidity divergences, which must be regularized. Thus beside the usual renormalization scale $\mu$, they also depend on a scale $\nu$ in the so-called rapidity regulator scheme introduced in~\cite{Chiu:2012ir}. At the same time, $S^{\text{global}}_{n{\bar n} n_J }$ is a wide-angle global soft function and $S^{cs}_{n_J}(\vec{k}_{4\perp}, R)$ is the collinear-soft function, with $n$, ${\bar n}$ and $n_J$ unit light-like vectors pointing along the $z$, $-z$ and the jet axis directions, respectively. For details about these soft functions and their perturbative expressions at the next-to-leading order, 
see~\cite{Buffing:2018ggv, Chien:2019gyf}. 

Let us now discuss the jet function $J_{c}(p_{JT} R, \mu)$, which encodes collinear radiations inside the jet. The NLO results for quark and gluon jet functions can be found in e.g.~\cite{Ellis:2010rwa,Liu:2012sz}. For completeness, the quark jet function $J_q$ for anti-$k_T$ algorithm is given by
\bea
J_q(p_{JT} R, \mu) &= 1+ \frac{\alpha_s}{\pi} C_F \left(L^2 - \frac{3}{2}L + \frac{13}{4} - \frac{3\pi^2}{8} \right),
\eea
where $L$ is the logarithm defined as
\bea
L = \ln\left(\frac{p_{JT} R}{\mu}\right). 
\eea
Thus the natural scale of the jet function is given by 
\bea
\mu_J \sim p_{JT}R.
\eea
At the same time, the jet function satisfies the renormalization group equation
\bea
\mu\frac{d}{d\mu} J_i(p_{JT} R, \mu) = \gamma_J^i(\mu) \, J_i(p_{JT} R, \mu),
\eea
which leads to the following solution
\bea
J_i(p_{JT} R, \mu) = J_i(p_{JT} R, \mu_J) \exp\left[\int_{\mu_J}^{\mu} \frac{d\mu'}{\mu'} \gamma_J^i(\mu')\right]\,,
\label{eq:J_RG}
\eea
with $i=q,g$ for quark and gluon jets. The anomalous dimensions $\gamma_J^i$ are given by
\bea
\gamma^i_J(\mu) = -2\,\Gamma^i_{\rm cusp}(\alpha_s) \,L + \gamma^i (\alpha_s),
\label{eq:gamma_J}
\eea
with $\Gamma^i_{\rm cusp}$ and $\gamma^i$ the cusp and non-cusp anomalous dimensions. They have the perturbative expansions $\Gamma^i_{\rm cusp} = \sum_n \Gamma_{n-1}^i \left(\frac{\alpha_s}{4\pi}\right)^n$ and $\gamma^i = \sum_n \gamma_{n-1}^i \left(\frac{\alpha_s}{4\pi}\right)^n$~\cite{Becher:2006mr,Becher:2009th,Jain:2011xz,Echevarria:2012pw,Moch:2004pa}. For example, 
\bea
& \Gamma_0^q = 4 C_F, \qquad  \gamma^q_0 = 6C_F,
\\
& \Gamma_{0}^g = 4 C_A, \qquad  \gamma^g_0 = 2\beta_0,
\eea
where $\beta_0 = \frac{11}{3} C_A - \frac{4}{3} T_F n_f$, with $T_F = \frac{1}{2}$ and $n_f$ the number of active quark flavors.

\subsection{Hadron distribution inside $Z$-tagged jets: $z_h$-dependence}
Now if we measure the longitudinal (along the jet direction) $z_h$ distribution of hadrons inside the $Z$-tagged jet, the factorized formalism can be written as
\bea
\frac{d\sigma^h}{d{\cal PS} \,dz_h} = & 
\sum_{a,b,c} \int d \phi_J \, 
\int \prod_{i=1}^4 d^2{\bm k}_{iT}  \delta^{2}({\bm q}_T - \sum_i^4 {\bm k}_{iT})
 f_a(x_a,k_{1T}^2, \mu, \nu) f_b(x_b,k_{2T}^2, \mu, \nu) 
 \nonumber \\
 &
  \times   
  S^{\text{global}}_{n{\bar n} n_J }({\bm k}_{3T}, \mu, \nu) 
  S^{cs}_{n_J}({\bm k}_{4T},R, \mu) 
  H_{a b\rightarrow c Z}(p_T, m_Z, \mu)  \, {\cal G}_{c}^{h}(z_h, p_{JT} R, \mu) \,,
\label{eq:h_z}
\eea
where we replace the jet function $J_{c}(p_{JT} R, \mu)$ in Eq.~\eqref{eq:jet} by the fragmenting jet function ${\cal G}_{c}^{h}(z_h, p_{JT} R, \mu)$~\cite{Chien:2015ctp,Jain:2011xz}. Here $z_h = p_h^+/p_J^+$, with $p_h^+$ and $p_J^+$ the large light-cone component of the hadron and the jet, respectively. The fragmenting jet function ${\cal G}_i^h(z_h, p_{JT} R, \mu)$ will no longer be purely perturbative since it involves the hadron in the jet, which is non-perturbative. However, ${\cal G}_i^h(z_h, p_{JT} R, \mu)$ can be matched onto the standard collinear fragmentation functions (FFs)~$D_{h/i}(z_h, \mu)$, 
\bea
{\cal G}_i^h(z_h, p_{JT} R, \mu) = \sum_j \int_{z_h}^1 \frac{dz}{z}{\cal J}_{ij}(z, p_{JT} R, \mu)\,D_{h/j}\left(\frac{z_h}{z}, \mu\right),
\eea
where one can find the coefficients ${\cal J}_{ij}$ at NLO in \cite{Chien:2015ctp,Waalewijn:2012sv}. For later convenience, let us reproduce the expression for ${\mathcal J}_{qq}$ here,
\bea
{\mathcal J}_{qq}(z, p_{JT}R, \mu) &= \delta(1-z) +  \frac{\alpha_s}{\pi} C_F \left[\delta(1-z) \left(L^2 - \frac{\pi^2}{24}\right) 
+ \frac{1+z^2}{(1-z)_+}\left(L+ \ln z\right) + \frac{1-z}{2} +(1+z^2)\left(\frac{\ln(1-z)}{1-z}\right)_+ \right]\,.
\label{eq:Jqq}
\eea
At the same time, it is important to realize that ${\cal G}_{i}^{h}(z_h, p_{JT} R, \mu)$ follows the same renormalization group equation as the jet function $J_{i}(p_{JT} R, \mu)$ in Eq.~\eqref{eq:gamma_J}, 
\bea
\mu\frac{d}{d\mu} {\cal G}_i^h(z_h, p_{JT} R, \mu) = \gamma_J^i(\mu) \,{\cal G}_i^h(z_h, p_{JT} R, \mu)\,,
\eea
which would evolve ${\cal G}_i^h$ from its natural scale, again $\mu_J\sim p_{JT}R$, up to the hard scale $\mu$ as
\bea
{\cal G}_i^h(z_h, p_{JT} R, \mu) = {\cal G}_i^h(z_h, p_{JT} R, \mu_J) \exp\left[\int_{\mu_J}^{\mu} \frac{d\mu'}{\mu'} \gamma_J^i(\mu')\right]\,.
\label{eq:G_RG}
\eea

\subsection{Hadron distribution inside $Z$-tagged jets: $j_\perp$-dependence}
Finally if we measure both the longitudinal $z_h$ and transverse momentum $j_\perp$ distribution of hadrons inside the $Z$-tagged jet, the factorized formalism can be written as
\bea
\frac{d\sigma^h}{d{\cal PS} \,dz_h\, d^2{\bm j}_\perp} = & 
\sum_{a,b,c} \int d \phi_J \, 
\int \prod_{i=1}^4 d^2{\bm k}_{iT}  \delta^{2}({\bm q}_T - \sum_i^4 {\bm k}_{iT})
 f_a(x_a,k_{1T}^2, \mu, \nu) f_b(x_b,k_{2T}^2, \mu, \nu) 
 \nonumber \\
 &
  \times   
  S^{\text{global}}_{n{\bar n} n_J }({\bm k}_{3T}, \mu, \nu) 
  S^{cs}_{n_J}({\bm k}_{4T},R, \mu) 
  H_{a b\rightarrow c Z}(p_T, m_Z, \mu)  \, {\cal G}_{c}^{h}(z_h, p_{JT} R, {\bm j}_\perp,\mu) \,,
\label{eq:h_jt}
\eea
where this time we have a TMD fragmenting jet function ${\cal G}_{c}^{h}(z_h, p_{JT} R, {\bm j}_\perp,\mu)$, and ${\bm j}_\perp$ is the transverse component of the hadron momentum with respect to the jet direction. We are interested in the small $j_\perp$ region, $j_\perp\ll p_{JT}R$, where ${\cal G}_{c}^{h}(z_h, p_{JT} R, {\bm j}_\perp,\mu)$ receives contributions from both collinear, and collinear-soft modes~\cite{Bain:2016rrv}. It can be further factorized as~\cite{Bain:2016rrv,Kang:2017glf}
\bea
{\cal G}_{i}^{h}(z_h, p_{JT} R, {\bm j}_\perp,\mu) = \int d^2{\bm k}_\perp d^2{\bm \lambda}_\perp\delta^2\left(z_h{\bm \lambda}_\perp + {\bm k}_\perp - {\bm j}_\perp\right) D_{h/i}(z_h, {\bm k}_\perp, \mu, \nu) S_{i}({\bm \lambda}_\perp, \mu, \nu R), 
\eea
where the collinear mode is described by the usual TMD FFs $D_{h/i}(z_h, {\bm k}_\perp, \mu, \nu)$, and the collinear-soft mode is captured by the soft function $S_{i}({\bm \lambda}_\perp, \mu, \nu R)$. Besides the usual renormalization scale $\mu$, the scale $\nu$ is again associated with the rapidity divergence. Here it might be instructive to point out the difference between the above refactorization and those for TMD hadron distribution inside a single inclusive jet produced in proton-proton collisions, $p+p\to {\rm jet}+h+X$, in~\cite{Kang:2017glf}, where an additional hard factor arises that captures out-of-jet radiation with characteristic scale $\sim p_{JT}R$. Here since we are studying $Z$+jet production in the back-to-back region, such out-of-jet radiation is not allowed at leading-power. This is because any out-of-jet radiation would generate $Z$-jet imbalance of the order $p_{JT}R\gg q_T$, which would thus move $Z$-jet away from the back-to-back configuration. 

Following the usual wisdom in TMD physics, we transform the above expression in the transverse momentum space into the coordinate ${\bm b}$-space as follows
\bea
{\cal G}_{i}^{h}(z_h, p_{JT} R, {\bm j}_\perp,\mu) = \int \frac{d^2{\bm b}}{(2\pi)^2} \, e^{i{\bm j}_\perp\cdot {\bm b}/z_h} D_{h/i}(z_h, {\bm b}, \mu, \nu) S_i({\bm b}, \mu, \nu R),
\label{eq:Gjt}
\eea
where the Fourier transform is defined as follows
\bea
D_{h/i}(z_h, {\bm b}, \mu, \nu) =& \frac{1}{z_h^2} \int d^2{\bm k}_\perp e^{-i{\bm k}_\perp\cdot{\bm b}/z_h} D_{h/i}(z_h, {\bm k}_\perp, \mu, \nu)\,,
\\
S_i({\bm b}, \mu, \nu R) =& \int d^2{\bm \lambda}_\perp e^{-i{\bm \lambda}_\perp\cdot {\bm b}} S_i({\bm \lambda}_\perp, \mu, \nu R)\,.
\eea
The perturbative results up to next-to-leading order and the renormalization for both $D_{h/i}(z_h, {\bm b}, \mu, \nu)$ and $S_i({\bm b}, \mu, \nu R)$ have been carefully studied in \cite{Kang:2017glf}. Over there we define the ``proper'' in-jet TMD fragmentation function ${\cal D}^{R}_{h/i}$ as
\bea
{\cal D}^{R}_{h/i}(z_h, {\bm b}, \mu) = D_{h/i}(z_h, {\bm b}, \mu, \nu) S_i({\bm b}, \mu, \nu R)\,,
\eea
where the rapidity divergence cancels between $D_{h/i}(z_h, {\bm b}, \mu, \nu)$ and $S_i({\bm b}, \mu, \nu R)$, and thus there is no rapidity divergence and thus no $\nu$-dependence on the left-hand side. We also find that ${\cal D}^{R}_{h/i}$ evolves as follows
\bea
{\cal D}^{R}_{h/i}(z_h, {\bm b}, \mu) =& \hat {\mathcal D}_{h/i}(z_h, {\bm b}, \mu_J) \exp\left[\int_{\mu_J}^{\mu}\frac{d\mu'}{\mu'}\left(-2\Gamma_{\rm cusp}^i(\alpha_s) L + \gamma^i(\alpha_s)\right)\right]\,,
\nnu
=& \hat {\mathcal D}_{h/i}(z_h, {\bm b}, \mu_J)  \exp\left[\int_{\mu_J}^{\mu} \frac{d\mu'}{d\mu'} \gamma_J^i(\mu')\right]\,.
\eea
where the equation holds when $\mu_J = p_{JT}R$, and $\hat {\mathcal D}_{h/i}(z_h, {\bm b}, \mu_J)$ are the ``properly''-defined TMD FFs, i.e., those measured in semi-inclusive deep inelastic scattering and/or back-to-back hadron pair production in $e^+e^-$ collisions~\cite{Collins:2011zzd}. Plug this result into Eq.~\eqref{eq:Gjt}, we obtain
\bea
{\cal G}_{i}^{h}(z_h, p_{JT} R, {\bm j}_\perp,\mu) =& \left[\int \frac{d^2{\bm b}}{(2\pi)^2} \, e^{i{\bm j}_\perp\cdot {\bm b}/z_h} \hat {\mathcal D}_{h/i}(z_h, {\bm b}, \mu_J)\right] \exp\left[\int_{\mu_J}^{\mu} \frac{d\mu'}{d\mu'} \gamma_J^i(\mu')\right]\,,
\nnu
\equiv &\hat {\mathcal D}_{h/i}(z_h, {\bm j}_\perp, \mu_J) \exp\left[\int_{\mu_J}^{\mu} \frac{d\mu'}{d\mu'} \gamma_J^i(\mu')\right]\,.
\label{eq:GTMD}
\eea
One of the most important observations is that the evolution factor, i.e., the exponential part on the right-hand side is the same for the jet function $J_i(p_{JT} R, \mu)$ in Eq.~\eqref{eq:J_RG}, the fragmenting jet function ${\cal G}_i^h(z_h, p_{JT} R, \mu)$ in Eq.~\eqref{eq:G_RG}, and the TMD fragmenting jet function ${\cal G}_{i}^{h}(z_h, p_{JT} R, {\bm j}_\perp,\mu)$ in Eq.~\eqref{eq:GTMD}. In other words, the renormalization group equation is the same for all of them. This is consistent with the factorized formalism, since the rest of the factors are the same for all three cases in Eqs.~\eqref{eq:jet}, \eqref{eq:h_z}, and \eqref{eq:h_jt}. This factor is different from the hadron distribution inside jets for single inclusive jet production, as extensively studied in e.g. Refs.~\cite{Kang:2017yde,Kang:2016ehg,Kang:2017glf,Kang:2017mda,Cal:2019hjc}. For single inclusive jet production, the renormalization group equations for the relevant jet functions follow time-like DGLAP equations. 

For the proper TMD fragmentation functions $\hat {\mathcal D}_{h/i}(z_h, {\bm j}_\perp, \mu_J)$, we use the same parametrization as in \cite{Kang:2017glf},
\bea
\hat {\mathcal D}_{h/i}(z_h, {\bm j}_\perp; \mu_J) &= \frac{1}{z_h^2}\int \frac{db}{2\pi} bJ_0(j_\perp b/z_h) C_{j\leftarrow i}\otimes D_{h/j}(z_h, \mu_{b_*}) e^{-S_{\rm pert}^i(b_*, \,\mu_J) - S_{\rm NP}^i(b, \,\mu_J)}\,,
\label{eq:Djt}
\eea
where we have used so-called $b_*$-prescription to avoid Landau pole of strong coupling $\alpha_s$~\cite{Collins:1984kg}, $C_{j\leftarrow i}$ are the coefficient functions, $S_{\rm pert}^i(b_*, \mu_J)$ is the perturbative Sudakov factor, and 
$S_{\rm NP}^i(b, \mu_J)$ is the non-perturbative Sudakov factor. Their expressions are all given in \cite{Kang:2017glf}, where TMD FFs are computed at next-to-leading order for $C_{j\leftarrow i}$ and at next-to-leading logarithmic level for $S_{\rm pert}^i(b_*, \mu_J)$. The integration in Eq.~\eqref{eq:Djt} involves Bessel function $J_0$ which is oscillating and we thus have used an optimized Ogata quadrature method developed in~\cite{Kang:2019ctl} to handle the integration for better numerical convergence and reliability. 

\section{Phenomenology at the LHC}
\label{sec:pheno}
In this section, we present numerical results for hadron distribution inside $Z$-tagged jets in proton-proton collisions and compare to the experimental measurements by the LHCb collaboration at the LHC. 

The LHCb collaboration has performed measurements for hadron distribution inside $Z$-tagged jets in proton-proton collisions at the center-of-mass energy $\sqrt{s} = 8$ TeV in the forward rapidity regions at the LHC. The jet rapidity is integrated over $2.5 < \eta_J < 4.0$, while the $Z$-boson rapidity is integrated over $2.0 < \eta_Z < 4.5$. The jets are reconstructed using the anti-$k_T$ algorithm with a jet size parameter of $R=0.5$~\cite{Aaij:2019ctd}. For the longitudinal distribution of hadrons inside jets, we define the jet fragmentation function as
\bea
F(z_h) = \left. \frac{d\sigma^h}{d{\cal PS} \, dz_h} \right/ \frac{d\sigma}{d{\cal PS} }\,,
\eea
where the numerator and the denominator are given by Eqs.~\eqref{eq:h_z} and \eqref{eq:jet}, respectively, and we have suppressed the dependence on the rapidity and transverse momentum for both the $Z$-boson and the jet in $F(z_h)$. At the same time, for the $j_\perp$-dependence of the hadrons inside the jet, we define
\bea
F(z_h, j_\perp) = \left. \frac{d\sigma^h}{d{\cal PS} \, dz_h dj_\perp} \right/ \frac{d\sigma}{d{\cal PS} }\,.
\eea
Note that the numerator can be easily computed from Eq.~\eqref{eq:h_jt}, with the azimuthal angle of ${\bm j}_\perp$ integrated over, and further multiplied by a factor of $j_\perp$. In the numerical computations, we use NLO DSS fragmentation functions for charged hadrons from~\cite{deFlorian:2007aj}. Other fragmentation functions such as NNFF1.1~\cite{Bertone:2018ecm} give similar results. 

\bef
\includegraphics[width=0.95\columnwidth]{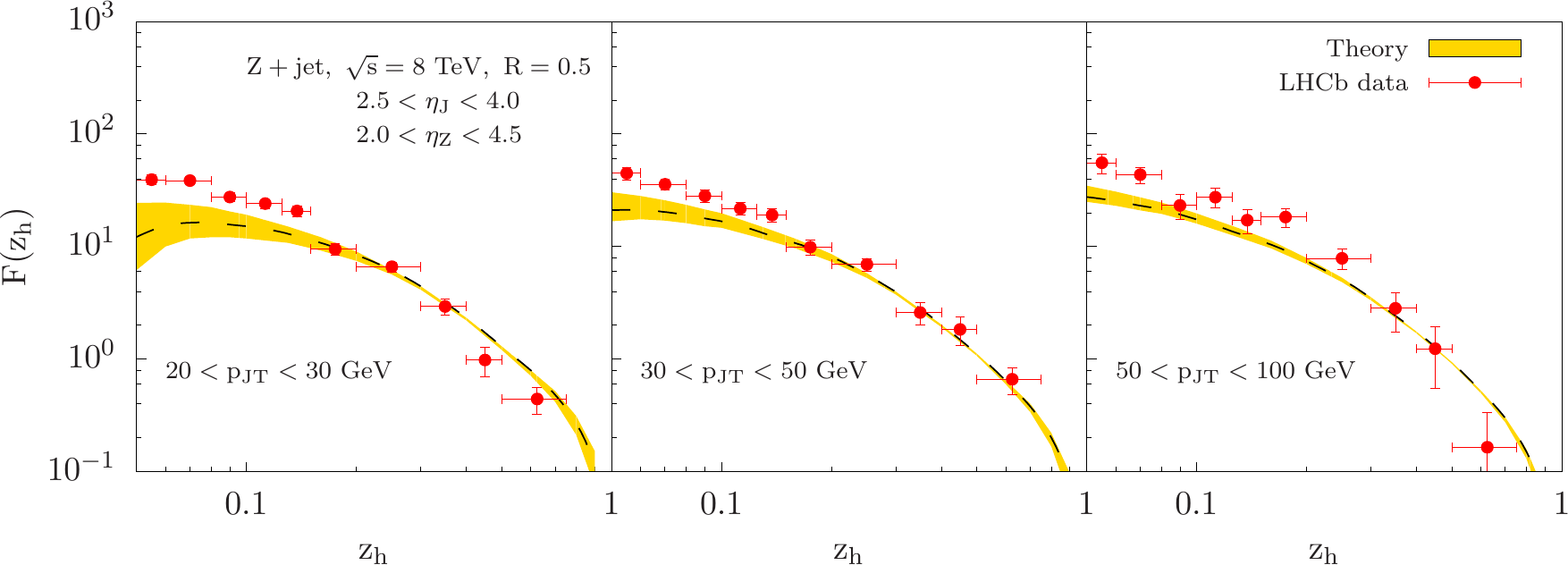}
\caption{Hadron distributions inside $Z$-tagged jets $F(z_h)$ in Eq.~\eqref{eq:h_z} are plotted as functions of $z_h$. From left to right, the three panels correspond to different jet transverse momenta: $20 < p_{JT} < 30$ GeV, $30 < p_{JT} < 50$ GeV, and $50 < p_{JT} < 100$ GeV. The yellow band is the theoretical uncertainty from the scale variation as explained in the text. The red solid data points are from LHCb collaboration~\cite{Aaij:2019ctd}.}
\label{fig:zh-diff}
\eef

In Fig.~\ref{fig:zh-diff}, we plot $F(z_h)$ as a function of $z_h$. We make the default scale choices of $\mu=\sqrt{p_T^2+m_Z^2}$ and $\mu_J=p_{JT}R$. We explore the scale uncertainty by varying $\mu$ and $\mu_J$ independently by a factor of two around their default values and by taking the envelope of these variations. From left to right, the three panels correspond to different jet transverse momenta: $20 < p_{JT} < 30$ GeV (left), $30 < p_{JT} < 50$ GeV (middle), and $50 < p_{JT} < 100$ GeV (right). We find that for the intermediate $0.1\lesssim z_h \lesssim 0.5$, our results describe the LHCb data reasonably well. However, when $z_h$ is either very small ($z_h\ll 1$) or very large ($z_h\to 1$), the description becomes worse. This is easily understood. From Eqs.~\eqref{eq:h_z} and~\eqref{eq:Jqq}, the coefficient functions such as ${\cal J}_{qq}$ contains $\ln z$ and $\left(\frac{\ln(1-z)}{1-z}\right)_+$, which become important for $z\ll 1$ and $z\to 1$, respectively. Thus one has to resum such types of logarithms: one might follow \cite{Anderle:2016czy} for $\ln z$ resummation, while for large-$z$ one could get insights from~\cite{Dai:2017dpc}. We leave such studies for future publication.

\bef
\includegraphics[width=0.95\columnwidth]{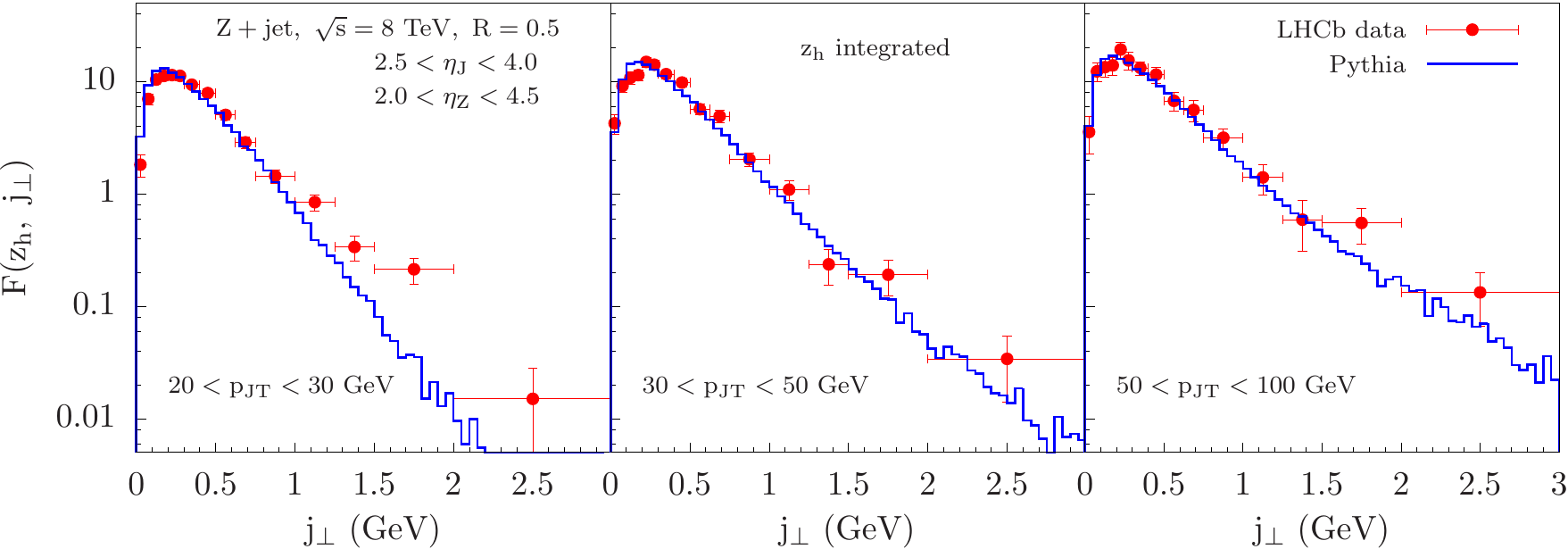}
\caption{The comparison between the LHCb data (red solid points) and the Pythia simulation (blue histogram) for hadron $j_\perp$ distribution. We integrate over the entire $z_h$ range.}
\label{fig:jt-allzh}
\eef

For the $j_\perp$-distribution of hadrons inside $Z$-tagged jets, LHCb formally integrates over the entire $0< z_h < 1$ region.~\footnote{There is a lower cut at a very small $z_h$, since LHCb only selects hadrons with $p_T^h > 0.25$ GeV.} From Eq.~\eqref{eq:Djt}, this would require that we know well the standard collinear fragmentation function $D_{h/i}(z_h, \mu)$ for the entire $0< z_h < 1$ region. However, typical global analysis for fragmentation functions only constrains the fragmentation functions for $z_h\gtrsim 0.05$. This fact thus hinders a more direct and transparent comparison between our theoretical calculations and the LHCb data, as we have observed previously~\cite{Kang:2017glf} for hadron distribution in inclusive jet production. To help the situation, in Fig.~\ref{fig:jt-allzh} we make a comparison between the LHCb data and the Pythia 8 simulation~\cite{Sjostrand:2007gs}. In the Pythia simulation, we make the same cuts as in the experiments and integrate over the entire $z_h$ range. As one can see clearly from Fig.~\ref{fig:jt-allzh}, the Pythia simulation gives a good description for the hadron $j_{\perp}$-distribution in the small and intermediate region. 

\bef
\includegraphics[width=0.95\columnwidth]{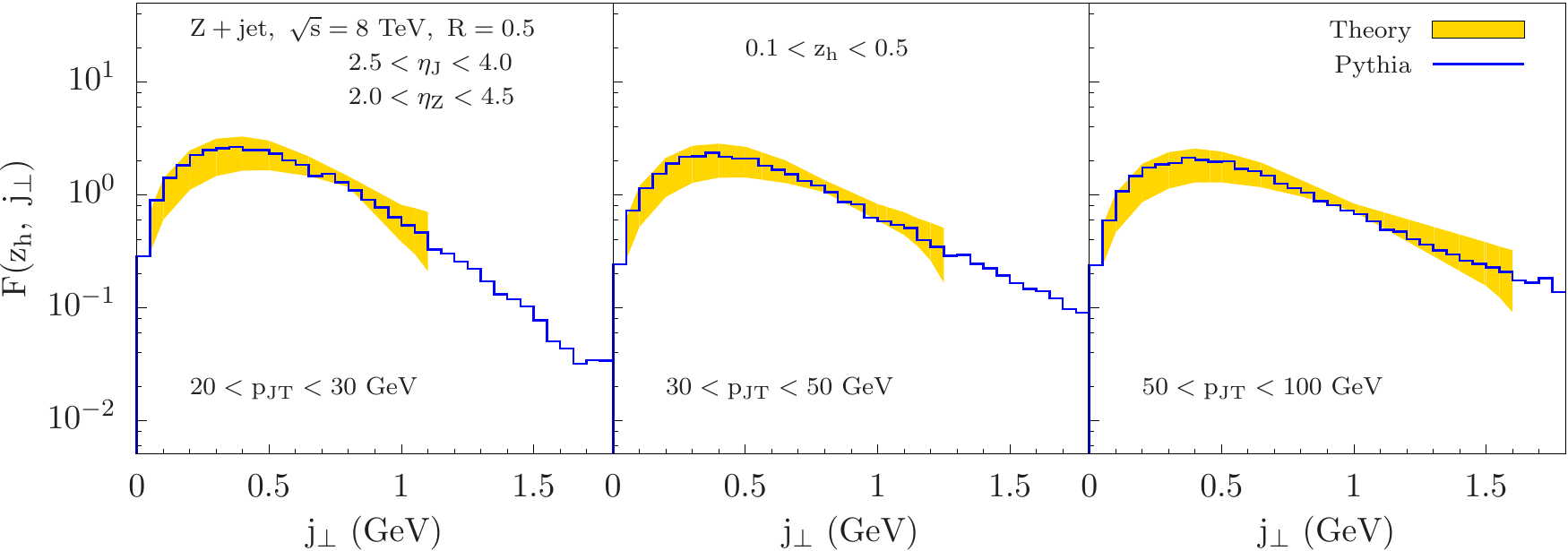}
\caption{The comparison between our theoretical computations (yellow bands) and the Pythia simulation (blue histogram) for hadron $j_\perp$ distribution. We integrate $z_h$ over the range $0.1 < z_h < 0.5$.}
\label{fig:jt-zh15}
\eef

Since Pythia simulations give such good descriptions of the LHCb data on hadron $j_\perp$-dependence, we thus could use Pythia 8 to simulate the hadron $j_\perp$-dependence, integrated for an appropriate $z_h$ range, which is suitable for comparison with our theoretical results. With this in mind, we perform such Pythia simulations and integrate over $0.1 < z_h < 0.5$. The simulations are presented in blue histograms in Fig.~\ref{fig:jt-zh15}. At the same time, we present our theoretical computations as yellow bands, which are generated the same as in Fig.~\ref{fig:zh-diff}, i.e., from the scale variation of $\mu$ and $\mu_J$ from their corresponding natural scales. We find that our TMD calculations agree well with the Pythia simulations. Note that our factorized formalism works only for the small $j_\perp \ll p_{JT}R$ region. For the relatively large $j_\perp$ region, one expects the so-called $Y$-term~\cite{Collins:2011zzd} to become important and has to be included to describe the data. This is why our theoretical curves stop at certain $j_\perp$ values. 

\section{Conclusion}
\label{sec:summary}
We study back-to-back $Z$-jet production in proton-proton collisions at the LHC. In particular, we concentrate on the longitudinal $z_h$ and transverse momentum $j_\perp$ distribution of hadrons inside $Z$-tagged jets. We find that the $z_h$-dependence is sensitive to the standard collinear fragmentation functions, while the $j_\perp$-dependence probes the transverse momentum dependent fragmentation functions (TMD FFs). The numerical calculations based on our theoretical formalism give good descriptions of the LHCb data for intermediate $z_h$ region. For $j_\perp$-dependence, since the experimental data are integrated over the entire $0<z_h<1$ region, the direct comparison is nontrivial if not impossible. For integrating over the intermediate $0.1 < z_h < 0.5$ region, our results agree well with the Pythia simulations for the relatively small $j_\perp$ region. For future measurements, we suggest to set up the binning in both $z_h$ and $j_\perp$, as this would lead to a more direct probing of TMD FFs. We expect our work to have important applications in studying fragmentation functions in vector-boson-tagged jet production in both proton-proton and nucleus-nucleus collisions. 

\section*{Acknowledgments}
We thank Christine Aidala and Joe Osborn for discussions and the correspondence on the LHCb experimental data. Z.K. is supported by the National Science Foundation under Grant No.~PHY-1720486. K.L. is supported by the National Science Foundation under Grants No.~PHY-1316617 and No.~PHY-1620628. J.T. is supported by the NSF Graduate Research Fellowships Program. H.X. is supported by NSFC of China under Project No.~11435004 and research startup funding at SCNU. This research is also supported within the framework of the TMD Topical Collaboration.

\bibliography{refs}
\bibliographystyle{h-physrev5}

\end{document}